\documentclass{Interspeech2024}
\usepackage{url}
\usepackage{multirow}

\interspeechcameraready

\title{BTS: Bridging Text and Sound Modalities for Metadata-Aided Respiratory Sound Classification}
\name{June-Woo}{Kim$^{*1,2}$} %June-Woo Kim - 1st author
\name{Miika}{Toikkanen$^{2}$} %Miika Toikkanen - 2nd author
\name{Yera}{Choi$^{3}$} %Yera Choi - 3rd author
\name{Seoung-Eun}{Moon$^{3\dagger}$} %Seong-Eun Moon - co-corresponding authors
\name{Ho-Young}{Jung$^{1\dagger}$} %Ho-Young Jung - co-corresponding authors
\address{
  $^1$Department of AI, Kyungpook National University, South Korea \\
  $^2$RSC LAB, MODULABS, South Korea \\
  $^3$NAVER Digital Healthcare LAB, NAVER AI LAB, NAVER Cloud, South Korea\thanks{\hspace{-1.7em}*\,work done while intern at NAVER Cloud \, $^{\dagger}$corresponding authors %\\ This research was supported by the MSIT (Ministry of Science and ICT), Korea, under the ITRC (Information Technology Research Center) support program (IITP-2024-2020-0-01808) supervised by the IITP (Institute of Information \& Communications Technology Planning \& Evaluation) and by Brian Impact Foundation, a non-profit organization dedicated to the advancement of science and technology for all.
  \\\hspace*{0em}Code is available at~\url{https://github.com/kaen2891/bts}.
  }
  }
\email{\{kaen2891, hoyjung\}@knu.ac.kr \quad seongeun.moon@navercorp.com}


\begin{document}

\maketitle


\begin{abstract} % no more than 1000 chars --> now 999 words
Respiratory sound classification (RSC) is challenging due to varied acoustic signatures, primarily influenced by patient demographics and recording environments.
To address this issue, we introduce a text-audio multimodal model that utilizes metadata of respiratory sounds, which provides useful complementary information for RSC.
Specifically, we fine-tune a pretrained text-audio multimodal model using free-text descriptions derived from the sound samples' metadata which includes the gender and age of patients, type of recording devices, and recording location on the patient's body.
Our method achieves state-of-the-art performance on the ICBHI dataset, surpassing the previous best result by a notable margin of 1.17\%.
This result validates the effectiveness of leveraging metadata and respiratory sound samples in enhancing RSC performance. Additionally, we investigate the model performance in the case where metadata is partially unavailable, which may occur in real-world clinical setting.
\end{abstract}

\section{Introduction}

Identifying abnormal respiratory sounds is pivotal for diagnosing and providing timely interventions for respiratory conditions. 
Automated detection of abnormal respiratory sounds has great potential to improve health and quality of life for those affected by respiratory diseases by identifying risks early and expediting first aid for potentially life-threatening conditions, such as pneumonia or chronic obstructive pulmonary disease.
% Issues from data scarcity
Machine learning approaches have been regarded as a promising way for automated detection of abnormal respiratory sounds.
Recently, a number of studies~\cite{kim2023stethoscope, kim2024repaugment, kim2023adversarial, bae23b_interspeech, moummad2023pretraining, nguyen2022lung, 9871440, wang2022domain, Gairola21} have tackled the respiratory sound classification (RSC) task and notably increased the performance by utilizing models that have been pretrained on large non-medical datasets~\cite{deng2009imagenet, audioset}, and then fine-tuned on a respiratory sound dataset~\cite{rocha2018alpha}.

%, which resulted in a notable performance enhancement for RSC. 
%, have been shown to offer a robust starting point that transfers well to respiratory sound data~\cite{kim2023stethoscope, kim2023adversarial, bae23b_interspeech}.

% Despite training reliable machine learning models that usually require extensive data, however, the scarcity of data remains a well-known challenge in the medical domain.
%Recent advancements in numerous neural network-based respiratory sound classification studies have increasingly focused on leveraging pretrained models to mitigate the challenges posed by the scarcity of applicable respiratory sound data \cite{kim2023stethoscope, kim2023adversarial, bae23b_interspeech, moummad2023pretraining, nguyen2022lung, 9871440, wang2022domain, Gairola21}. 
%Such models, trained on large and diverse datasets~\cite{deng2009imagenet, audioset}, have been shown to offer a robust starting point that transfers well to respiratory sound data~\cite{kim2023stethoscope, kim2023adversarial, bae23b_interspeech}. 

Nevertheless, the inherent heterogeneity of respiratory sound data presents an obstacle to further performance improvement in RSC.
The heterogeneity arises from differences in patient demographics, recording devices, and environmental conditions, which can significantly impact the acoustic properties of respiratory sounds~\cite{kim2023stethoscope}. This may lead to poor generalization on unseen data, particularly in cases underrepresented by the training data.
%minority classes of metadata.
ICBHI~\cite{rocha2018alpha}, one of the widely adopted respiratory sound datasets, provides metadata that associates the recorded audio with attributes of patients and recording environments. 
Such metadata may be useful for addressing difficulties caused by heterogeneity.

Some previous work has adapted the metadata associated with respiratory sound for RSC to mitigate the heterogeneity issue.
For instance, incorporating demographic information of patients such as age and gender into the pretraining process provides better representations of respiratory audio samples~\cite{moummad2023pretraining}.
%particularly in small and imbalanced datasets
% due to the correlations  with presence of lung diseases?
Moreover, metadata concerning the recording environment (i.e., stethoscope) also provides useful information. SG-SCL~\cite{kim2023stethoscope} employed domain-transfer techniques to reduce the effect of heterogeneity by regarding different types of recording devices as distinct domains. 
Despite the potential benefits of leveraging the metadata, these previous works did not fully incorporate it as text data into the model inputs.
%these previous works did not fully incorporate it as text data into the model inputs.
%Some previous work has adapted metadata, but typically regard it as a peripheral component rather than incorporating it directly into the model. 
%\red{Despite the potential benefits of leveraging metadata information, these previous works did not attempt to fully incorporate the text data.}
% Among the works that utilize this metadata, SG-SCL~\cite{kim2023stethoscope} showed that cross-domain techniques and knowledge about the recording device can be used to align the features extracted with different stethoscopes and mitigate the distribution inconsistency, thereby reducing the performance hit caused by variations in the recording device acoustics.
%demonstrated that the cross-domain between different recording devices can reduce the distribution inconsistency issue and lead the improved performance. 
%In other words, this study exploited available information about the recording device during training in the perspective of domain adaptation and aligned the features extracted from the respiratory audio with different stethoscopes to reduce the performance hit caused by variations in the recording device acoustics.
% Pretraining with metadata~\cite{moummad2023pretraining} demonstrated that incorporating demographic information such as age and gender can improve the feature representation learned from respiratory audio samples, particularly in small and imbalanced datasets. 

Recent developments in multimodal models, exemplified by Contrastive Language-Image Pretraining (CLIP)~\cite{radford2021learning} for text and image data and Contrastive Language-Audio Pretraining (CLAP)~\cite{elizalde2023clap, wu2023large} for text and audio data, offer a flexible framework for integrating text data with non-textual data.
% Previous research on language-image multimodal models has shown that integrating images with text inputs not only enhances the generation of more effective rationales~\cite{zhang2023multimodal}, but also improves the creation of textual descriptions from MRI scans~\cite{kwon2023large} or X-ray images~\cite{ji2024vision}, underscoring the value of combining multiple modalities in interpreting complex information.
% add multimodal learning of language and other bio signals
Several studies~\cite{hollenstein2021decoding, wang2022open, NEURIPS2023_1f2fd233, qiu2023can} have demonstrated the effectiveness of language-EEG multimodal models for the sentiment classification and EEG-to-text decoding tasks.
Recognizing the success of multimodal models and the demonstrated benefits of multimodal data in healthcare tasks, it is compelling to consider them for RSC, where such method has not yet been explored.
In this paper, we take a step into a new direction and fully make use of the respiratory audio metadata by adapting a text-audio multimodal model, aiming not only to leverage the metadata as an additional learning signal, but to benefit from the further context during the inference stage.
Building on the foundation of contrastive language-audio pretrained models, our work incorporates the respiratory audio metadata alongside the sound recordings. 
To this end, we format the patient's metadata into descriptions derived from key attributes including age, gender, recording device, and recording location on the body, and encode them with respiratory sound data into shared feature representation by the pretrained encoders. 
With these joint representations, we train a classification head for the RSC task.

% Results
\begin{figure*}[t!]
    % \vspace{-5mm}
    \centering
    \includegraphics[width=1.0\linewidth]{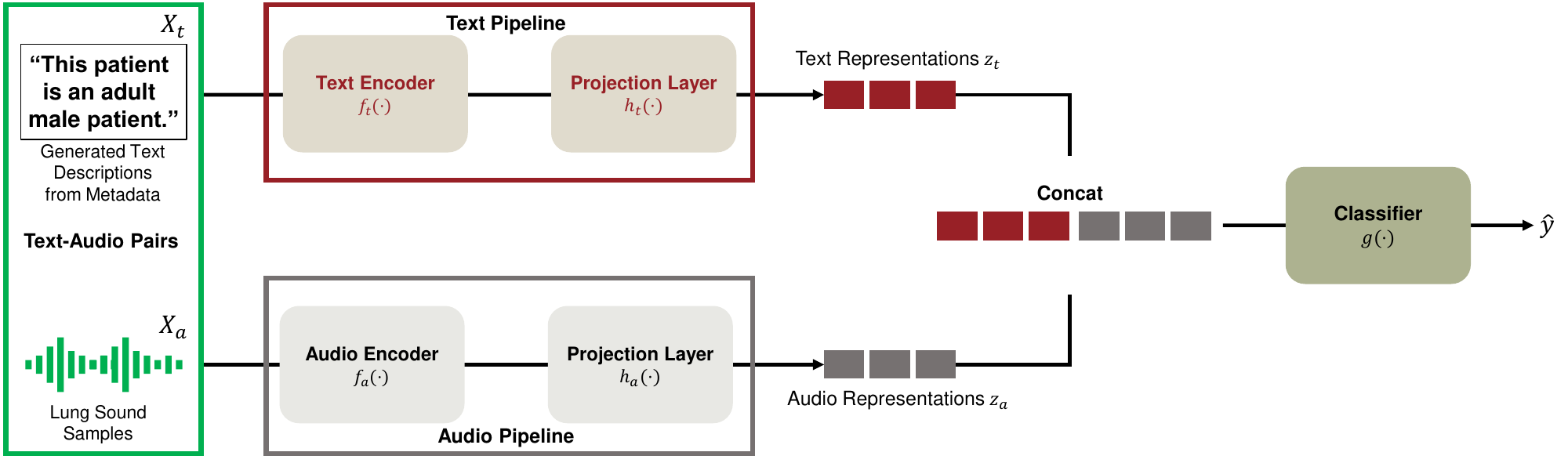}
    % \vspace{-20pt}
    \vspace{-5mm}
    \caption{An overall illustration of the proposed BTS architecture. The pretrained text and audio encoders extract feature representations of text description derived from metadata and respiratory sound samples, respectively. After the projection, the representations are integrated by a concatenation operation and used for RSC.}
    \label{fig:proposed}
\end{figure*}
Our approach, which we name the \textbf{\emph{BTS}} (\textbf{\emph{B}}ridging the \textbf{\emph{T}}ext and \textbf{\emph{S}}ound modalities), a method that leverages multimodal text-audio model to fully exploit the potential of respiratory audio metadata, achieves the state-of-the-art (SOTA) result on the ICBHI dataset, outperforming upon the previous best~\cite{bae23b_interspeech} by 1.17\%. 
%We also demonstrate that fine-tuning the pretrained audio encoder alone is sufficient to surpass the previously established state-of-the-art performance.
%We further present that simply fine-tuning with the pretrained audio encoder only can beat the previous state-of-the-art performance.
Our results reveal the capability of contrastive language-audio pretraining to improve RSC both in audio-only and multimodal settings.
Moreover, we demonstrate that our method retains its performance gains in the absence of metadata during the inference. 
This result suggests that our approach can be adopted for practical clinical settings where additional information other than audio signals may be unavailable.
%We find that our method mostly retains its performance gains in the absence of metadata and generalizes to variation in the content and format, providing flexibility in realistic clinical settings where information may be partially unavailable or stored in different formats. 
%Our main contributions can be summarized as follows: 
Our main contributions are as follows: 
\begin{itemize} 
% \item We demonstrate that single-modal from pretrained multimodal models can be a strong audio encoder for respiratory sound classification tasks.
\item We show that leveraging metadata of respiratory sounds improves the RSC performance.
%by employing multimodal models. 
Our approach sets the new SOTA performance on the ICBHI dataset.
\item We thoroughly explore ways to utilize the metadata considering a real clinical setting where the type of metadata differs from the expectation, or the metadata is partially or totally unavailable. We demonstrate that our method robustly performs RSC in such scenarios.
%robustly performs RSC with the partial absence of metadata, with an unseen type of metadata, and even without the metadata.
\item We analyze how the different types of metadata affect the performance of our model. Our result shows that the information about the recording environment, such as the type of recording stethoscopes and recording locations of the human body are particularly helpful in minimizing the effect of heterogeneity in respiratory sounds.
\end{itemize}

% Detecting abnormal respiratory sounds is crucial to providing first aid for potential life-threatening risks.
% 1. Presents the recent works \& trends. 

% i) Lack of medical data issue 

% Explain previous research. They are widely focused on using pretrained models for improved performance.

% ii) Heterogeneity of metadata factors: Patient demographics + Different metadata

% Explain about previous research on metadata: SCL for metadata paper and Cross-Domain + SCL paper

% 2. Tackle the issue (motivation)

% Why not use the metadata as an additional training index?

% In the NLP domain (LLM), several works made patient descriptions to handle the tasks (Patient descriptions generations. Generate Rationale. Totally ask LLM model.)

% 3. Our suggestion

% Inspired by this, we introduce a text-audio multimodal model that utilizes metadata of respiratory sounds, providing useful complementary information for RSC. Specifically, blah blah...

% 4. Results

% Our results acehives new SOTA

% Analysis of our results show, ...
\section{Method}

We introduce \textbf{\emph{B}}ridging \textbf{\emph{T}}ext and \textbf{\emph{S}}ound modalities (BTS), an approach that leverages multimodal text-audio model to fully exploit the potential of respiratory audio metadata. 
%By incorporating the metadata alongside the respiratory sound recordings, BTS can extract richer and more informative representations, ultimately achieving improved performance in RSC tasks.
To mitigate the heterogeneity of respiratory sounds, we propose to explicitly utilize the metadata, which we expect to capture the significant sources of acoustic variability. By integrating this metadata, we aim to reduce the heterogeneity issue and improve the RSC performance. Toward this goal, we propose the adoption of a multimodal text-audio model for RSC, as depicted in Figure~\ref{fig:proposed}.
%To mitigate the heterogeneity of respiratory sounds, we propose to explicitly utilize metadata, which we expect to imply the source of variability inherent in respiratory sounds. To this end, we advocate adoption of a multimodal text-audio model for RSC. The proposed method is illustrated in Figure~\ref{fig:proposed}.

\subsection{CLAP Model}
While the metadata of respiratory sounds can be employed for RSC in several different ways, a free text format is flexible and easily applicable to human-produced data such as medical records. 
%For instance, the metadata can be described by a categorical vector, where each element indicates different categories of the metadata. However, this approach requires a specific format and cannot handle missing values.
For instance, the metadata can be described by a vector of numeric values where each element indicates different metadata attributes. However, this approach is usually vulnerable to changes in input sources, such as missing data and unseen data types.
In contrast, encoders for free text data are trained to understand the given input, which makes approaches utilizing input data in a free text format robust to the changes.
%In contrast, encoders for free text data learn to grasp the meaning of the text, making methods that use free-text formats adaptable to changes and variations.
%For instance, the metadata can be described by a vector of numeric values where each value indicates different categories of the metadata. However, this approach is usually vulnerable to changes in input sources such as data missing and unseen data types.
%In contrast, recent encoders for free text data are trained to understand the given input, which makes approaches utilizing input data in a free text format robust to the changes.
For this reason, we use CLAP (Contrastive Language-Audio Pretraining)~\cite{wu2023large} as our starting point. The CLAP model includes both text and audio encoders, which are trained on the large-scale LAION-Audio-630K~\cite{wu2023large} dataset including diverse audio data.
%Accurately classifying respiratory sounds is a challenging task because of their inherent variability across humans.
%Specifically, variations in patient demographics, recording stethoscopes, and environmental factors can introduce significant inconsistencies in the respiratory sound samples, leading to limited generalization performance of the model.
%In that respect, a potential solution can be incorporating all available metadata as an additional input modality alongside the lung sound data.
%To achieve this, we propose employing a pretrained Contrastive Language-Audio Pretraining (CLAP)~\cite{wu2023large} model for fine-tuning, which can effectively handle both textual and audio data together.

%Given the text and audio data denoted as $X_{i}^{t}$ and $X_{i}^{a}$ (e.g., the waveform of a dog barking and `This is a sound of dog barking.') where $i \in[0, N]$ indicates the data index within a batch of size $N$, 
Given the text and audio data denoted as $X_{i}^{t}$ and $X_{i}^{a}$ where $i \in[1, N]$ indicates the data index within a batch of size $N$, 
% For each pair of text-audio data within a batch of size $N$, with each data point denoted as $X_{i}^{t}$ for text and $X_{i}^{a}$ for audio, where $i\ \in[0, N]$, 
CLAP processes the text and audio data independently through dedicated encoders $f_t(\cdot)$ and $f_a(\cdot)$ for each modality. %The embeddings $z_t \in \mathbb{R}^{N \times d}$ and $z_a \in \mathbb{R}^{N \times d}$ produced by the encoders are projected onto a $d$-dimensional shared embedding space through projection layers $h_t(\cdot)$ and $h_a(\cdot)$.
The embedding vectors produced by the encoders are projected onto a $d$-dimensional shared embedding space through projection layers $h_t(\cdot)$ and $h_a(\cdot)$.
% Given each pair of text-audio data within a batch of size $N$, represented by $X_{i}^{t}$ and $X_{i}^{a}$ for each data point $i\ \in[0, N]$, the CLAP model first processes them separately using the text and audio encoders, as denoted $f_t(\cdot)$ and $f_a(\cdot)$ respectively. The text-audio output representations are then fed into text and audio projection layers $h_t(\cdot)$ and $h_a(\cdot)$:
\begin{align}
  z_t &= h_t(f_t(X_{i}^{t})), \nonumber \\ 
  z_a &= h_a(f_a(X_{i}^{a})).
  \label{equ1}
\end{align}
The CLAP model is trained to maximize the similarity between the text and audio embeddings by contrasting them with negative samples (i.e., mismatched text or audio embeddings obtained from $X_{j\in[1, N]; j \ne i}^t$ or $X_{j\in[1, N]; j \ne i}^a$).
% The CLAP model is trained with the contrastive learning paradigm to maximize the similarity between the text embedding $z_t$ and the audio embedding $z_a$ using similarity matrix $C \in \mathbb{R}^{N \times N}$, where it contains $N$ correct pairs along the diagonal and $N^2 - N$ incorrect pairs elsewhere. 
% where $z_t \in \mathbb{R}^{N \times d}$, $z_a \in \mathbb{R}^{N \times d}$ and $d$ is joint multimodal space of dimension, respectively. The CLAP model is trained with the contrastive learning paradigm to maximize the similarity between the text embedding $z_t$ and the audio embedding $z_a$ using similarity matrix $C \in \mathbb{R}^{N \times N}$, where it contains $N$ correct pairs along the diagonal and $N^2 - N$ incorrect pairs elsewhere. 

\begin{table}[!t]
  \caption{Examples of generated text descriptions derived from metadata. `All' is the case that includes all attributes: age, sex, recording location, and recording device.}
  %\caption{Examples of generated text descriptions derived from metadata. `All' represents the case of adult, male, left anterior, and Meditron.}
  \label{table1_generated_descriptions}
  \centering
  \addtolength{\tabcolsep}{1pt}
    \resizebox{\linewidth}{!}{
  \begin{tabular}{ll}
    \toprule
    Metadata & Generated text descriptions \\
    \hline \midrule
    Age & This patient is an adult patient. \\
    Sex & This patient is a male patient. \\
    Loc & This sound was recorded from the left anterior chest. \\
    Dev & This sound was recorded with a Meditron stethoscope. \\
    \midrule
    Age-Loc-Dev & This sound was recorded from the left anterior chest \\ & of an adult patient, using a Meditron stethoscope. \\
    ......... & ......... \\
    \midrule
    All & This sound was recorded from the left anterior chest \\ & of an adult male patient, using a Meditron stethoscope. \\
    
    \bottomrule
  \end{tabular}
  
  \vspace{-5mm}
  }
\end{table}

\subsection{Text Description Generation for Metadata}
% \subsection{Generating Text Descriptions from Metadata}
Among the metadata available in the ICBHI~\cite{rocha2018alpha} dataset, we choose four types of data as follows: age (adult or pediatric) and gender (male or female) of patients, recording location on the chest of the patients (trachea, anterior left, anterior right, posterior left, posterior right, lateral left, or lateral right), and type of recording devices (Meditron, LittC2SE, Litt3200, or AKGC417L). Using the attributes, we construct simple text descriptions. A generated description can include any combination of the attributes, totaling 644 unique texts. Table~\ref{table1_generated_descriptions} illustrates a few examples with different combinations of metadata.
%that we deem important for RSC, that is,

% To incorporate metadata into the CLAP model, we automatically generate textual descriptions based on four data points: age group, gender, recording location, and stethoscope type.
% As shown in Table~\ref{table1_generated_descriptions}, the metadata can be chosen as a single or their combinations. 
%For instance, using all available metadata information would be generated as below: `This sound was recorded from the left anterior chest of an adult male patient, using a Meditron stethoscope.'.

\subsection{Bridging Text and Sound Modalities}
%\subsection{BTS: Bridging Text and Sound Modalities}

As shown in Figure~\ref{fig:proposed}, we train the text and audio encoders of CLAP for RSC by using the respiratory sound samples and generated text descriptions. For classification, we concatenate text and audio representations $z_t$ and $z_a$ from text and audio pipelines as described in Figure~\ref{fig:proposed}.
Consequently, we can obtain the multimodal combined representations $z = concat(z_t, z_a)$ where $z \in \mathbb{R}^{N \times 2d}$.
We then simply add a 4-dimensional linear layer for classifier $g(\cdot)$ followed by softmax function and train it with the Cross-Entropy loss $\mathcal{L}_{\text{CE}}$ (division by $N$ is omitted):
\vspace{-2mm}
\begin{align}
    \mathcal{L}_{\text{CE}} = -\sum_{i=1}^n\! \, y_{i}\! \, \log \, \!(\hat{y_{i}}),
\end{align}
where $n$ is number of samples, $y$ is the respiratory sound label $\in\{$normal, crackle, wheeze, both$\}$, and $\hat{y}$ is the predicted probabilities obtained by the classifier. %Note that our method is supervised classification task, not zero-shot classification.

\section{Experimental Setup}
\subsection{Dataset}

We utilized the ICBHI Respiratory dataset~\cite{rocha2018alpha}. The dataset contains a total of approximately 5.5 hours of respiratory sound recordings with pre-defined and balanced splits for training (60\%) and test (40\%) without patient overlap. There are 4,142 training and 2,756 testing respiratory cycles across four classes. Table~\ref{table2_dataset} illustrates the details of the ICBHI dataset. We binarize the age as the adult (over 18 years old) or pediatric (18 years old or under) for simplicity. Other than the age, we follow the metadata information of the official ICBHI records. Body mass index (BMI) data, which was provided only for adult patients, are solely employed for further analysis. For non-adults, we calculated it using their weight and height data.

%There are 4,142 training and 2,756 testing respiratory cycles across four classes: \emph{normal} (49.8\%--57.29\%), \emph{crackle} (29.30\%--23.55\%), \emph{wheeze} (12.10\%--13.97\%), and \emph{both} crackle and wheeze (8.80\%--5.19\%).
%The metadata used for in this study is as follows: \emph{age}$\ \in \{$adult, pediatric$\}$, \emph{sex}$\ \in \{$male, female$\}$, recording \emph{location}$\ \in \{$trachea, left anterior, right anterior, left posterior, right posterior, left lateral, right lateral$\}$, and recording \emph{stethoscope}$\ \in \{$Meditron, LittC2SE, Litt3200, AKGC417L$\}$. 
%Age is binarized, classifying patients as adults (over 18 years old) or pediatrics (non-adults). We followed other metadata information as official ICBHI records. For the ablation study, BMI was only available for adult patients; for non-adults, we calculated using their weight and height data.

%$i \in \{ normal, crackle, wheeze, both \}$

\subsection{Training Details}
Following the data pre-processing described in~\cite{kim2023stethoscope, kim2023adversarial, bae23b_interspeech, Gairola21}, we extracted the respiratory cycles from the waveform samples and standardized them to have a duration of 8 seconds. 
We then conducted resampling to 48kHz to match the pretraining data of CLAP. 
We employed the CLAP~\cite{wu2023large} model pretrained on the LAION-Audio-630K~\cite{wu2023large} dataset for all experiments. %\footnote{\url{https://huggingface.co/laion/clap-htsat-fused}}.
%both audio encoder only setting, as well as, BTS training.
The maximum length of the text descriptions is limited to 64 tokens, which was sufficient for avoiding truncation of text.
%For the BTS training, the maximum length of all generated descriptions derived from metadata was set to a fixed value of 64 tokens.
%We used the hidden size of
We fine-tuned the models using the Adam optimizer~\cite{kingma2014adam} with an initial learning rate of 5e--5. The learning rate was adjusted by cosine scheduling through a total of 50 epochs of training with a batch size of 8. To reduce the impact of random initialization, we conducted the experiments with five different random seeds.%\footnote{Note that even though the random seed is fixed, results can differ due to the non-determinism issue of PyTorch \emph{upsample-bicubic2d-backward} in the implementation of Swin Transformer module in CLAP.}. 

\subsection{Metrics}
We adapt the \emph{Specificity} ($S_{p}$), \emph{Sensitivity} ($S_{e}$), and their average ($Score$) as performance metrics for RSC, following the definitions in~\cite{rocha2018alpha}.
%These metrics provide a comprehensive view of both true positive and true negative rates, which are crucial for evaluating model performance in datasets like ICBHI, where the distribution of labeled classes is imbalanced as illustrated in Table~\ref{table2_dataset}.
All reported values of $S_p$, $S_e$, and $Score$ are the mean and variance from the five runs with different seeds.

% \subsection{\red{Audio Single Modal}}
\subsection{Baselines}
We compare the proposed method with previous studies including the current SOTA method~\cite{bae23b_interspeech}, which uses Audio Spectrogram Transformer (AST)~\cite{gong21b_interspeech} as a backbone model. 
We also consider the result based solely on the audio embedding of CLAP ($z_a$ in Equation~(\ref{equ1})) as an additional baseline, which we denote as Audio-CLAP.

\begin{table}[!t]
    \centering
    \caption{Details of the ICBHI dataset including the number of audio samples for each class and the types of metadata. L/R stands for left or right.}
    % \vspace{-1pt}
    \label{table2_dataset}
    \addtolength{\tabcolsep}{1pt}
    \resizebox{\linewidth}{!}{
    %\begin{tabular}{p{1pt}lccc}
    \begin{tabular}{clccc}
    \toprule
    & Label & Train & Test & Sum \\
    \hline \midrule
    \multirow{4}{*}{\multirow{4}{*}\textbf{Lung Sound}} & Normal & 2,063 & 1,579 & 3,642 \\
    & Crackle & 1,215 & 649 & 1,864 \\
    & Wheeze & 501 & 385 & 886 \\
    & Both & 363 & 143 & 506 \\
    \midrule

    & Type & \multicolumn{3}{c}{Metadata Label} \\
    \hline \midrule

    \multirow{5}{*}{\multirow{4}{*}\textbf{Metadata}} & Age & \multicolumn{3}{c}{Adult, Pediatric} \\
    & Sex & \multicolumn{3}{c}{Male, Female} \\
    & Location & \multicolumn{3}{c}{Trachea, L/R Anterior, L/R Posterior, L/R Lateral} \\
    & Stethoscope & \multicolumn{3}{c}{Meditron, LittC2SE, Litt3200, AKGC417L} \\
    & Others & \multicolumn{3}{c}{BMI (Adult only), Weight/Height (Pediatric only)} \\
    \bottomrule
    \end{tabular}}
\end{table}

%for the RSC task in addition to the proposed multimodal approach. 
% Previous studies on state-of-the-art and its comparable methods used Audio Spectrogram Transformer~\cite{gong21b_interspeech}, which is pretrained on both ImageNet~\cite{deng2009imagenet} and AudioSet~\cite{audioset}. Fine-tuning AST for the RSC task showed significant performance improvement compared with other approaches. One of the reasons for such success is larger datasets that the AST model was pretrained on. In this regard, the audio encoder of CLAP, which is also pretrained with the large-scale LAION-Audio-630K~\cite{wu2023large} dataset, is expected show the comparable performance in RSC.
% To compare recent promising results in the respiratory sound classification task, we use the audio encoder from the CLAP model. To this end, we only use the audio embedding $z_a$ in equation~(\ref{equ1}). As a result, respiratory sound samples $X_{i}^{t}$ are only fed into the audio encoder $f_a(\cdot)$ followed by audio projector $h_a(\cdot)$ in this setting.

\section{Results}
\label{section:results}

\subsection{Main Results} %Main % Comparison with Other Studies

\begin{table*}[!t]
    \centering
    %\vspace{-3mm}
    \caption{The RSC performance on the ICBHI dataset with the official 60--40\% train--test split. 
    Here, in the Pretraining Data column, IN, AS, and LA refer to ImageNet~\cite{deng2009imagenet}, AudioSet~\cite{audioset}, and LAION-Audio-630K~\cite{wu2023large}, respectively. $*$ denotes the previous state-of-the-art ICBHI Score.
    The \textbf{Best} and {\underline{second best}} results are highlighted by the bold characters and underlines.}
    
    \label{table3_main_results}
    \renewcommand{\arraystretch}{1}
    \addtolength{\tabcolsep}{8pt}
    \resizebox{\linewidth}{!}{
    \begin{tabular}{lccc|lll}
    \toprule
    Method & Backbone & Pretraining Data & Venue & $S_p$\,(\%) & $S_e$\,(\%) & \textbf{Score}\,(\%) \\
    \hline \midrule

    SE+SA \cite{yang2020adventitious} & ResNet18 & - & \textit{INTERSPEECH`20} & {81.25} & 17.84 & 49.55 \\

    LungRN+NL \cite{ma2020lungrn+} & ResNet-NL & - & \textit{INTERSPEECH`20} & 63.20 & 41.32 & 52.26 \\
    
    RespireNet \cite{Gairola21} (CBA+BRC+FT) & ResNet34 & IN & \textit{EMBC`21} & 72.30 & 40.10  & 56.20 \\

    Chang \textit{et al.} \cite{chang22h_interspeech} & CNN8-dilated & - & \textit{INTERSPEECH`22} & 69.92 & 35.85 & 52.89 \\
    
    Ren \textit{et al.} \cite{ren2022prototype} & CNN8-Pt & - & \textit{ICASSP`22} & 72.96 & 27.78 & 50.37 \\
    
    Wang \textit{et al.} \cite{wang2022domain} (Splice) & ResNeSt & IN & \textit{ICASSP`22} & 70.40 & 40.20 & 55.30 \\

    Late-Fusion \cite{9871440} & Inc-03\,+\,VGG14 & IN & \textit{EMBC`22} & \textbf{85.60} & 30.00 & 57.30 \\

    Nguyen \textit{et al.} \cite{nguyen2022lung}\,(StochNorm) & ResNet50 & IN & \textit{TBME`22} & 78.86 & 36.40 & 57.63 \\
    
    Nguyen \textit{et al.} \cite{nguyen2022lung}\,(CoTuning) & ResNet50 & IN & \textit{TBME`22} & 79.34 & 37.24 & $\text{58.29}$ \\

    Moummad \textit{et al.} \cite{moummad2023pretraining} & CNN6 & AS & \textit{WASPAA`23} & 70.09 & 40.39 & 55.24 \\
    
    Moummad \textit{et al.} \cite{moummad2023pretraining}\,(SCL) & CNN6 & AS & \textit{WASPAA`23} & 75.95 & 39.15 & 57.55 \\    
    
    Bae \textit{et al.} \cite{bae23b_interspeech}\, (Fine-tuning)  & AST & IN\,+\,AS & \textit{INTERSPEECH`23} & $\text{77.14}$ & $\text{41.97}$ & $\text{59.55}$ \\
    
    Bae \textit{et al.} \cite{bae23b_interspeech}\, (Patch-Mix CL) & AST & IN\,+\,AS & \textit{INTERSPEECH`23} & $\text{81.66}$ & $\text{{43.07}}$ & $\text{62.37}^\textbf{*}$ \\
    
    Kim \textit{et al.} \cite{kim2023adversarial}\, (AFT on Mixed-500) & AST & IN\,+\,AS & \textit{NeurIPSW`23} & $\text{{80.72}}$ & $\text{{42.86}}$ & $\text{61.79}$ \\
    
    Kim \textit{et al.} \cite{kim2023stethoscope}\, (SG-SCL) & AST & IN\,+\,AS & \textit{ICASSP`24} & $\text{{79.87}}$ & $\text{{43.55}}$ & $\text{{61.71}}$ \\

    Kim \textit{et al.} \cite{kim2024repaugment}\, (RepAugment) & AST & IN\,+\,AS & \textit{EMBC`24} & $\text{\underline{82.47}}$ & $\text{40.55}$ & $\text{61.51}$ \\

    \midrule

    \textbf{Audio-CLAP [ours]} & \text{CLAP} & LA & \textit{INTERSPEECH`24} & $\text{{80.85}}_{\pm 3.33}$ & $\text{\underline{44.67}}_{\pm 3.77}$ & $\text{\underline{62.56}}_{\pm 0.37}$ \\
    
    \textbf{BTS [ours]} & CLAP & LA & \textit{INTERSPEECH`24} & $\text{81.40}_{\pm 2.57}$ & $\text{\textbf{45.67}}_{\pm 2.66}$ & $\text{\textbf{63.54}}_{\pm 0.80}$ \\
    
    \bottomrule
    \end{tabular}}
    \vspace{-5mm}
\end{table*}

\begin{table}[!t]
    \centering
    %\vspace{-3mm}
    \caption{A comparison of the ICBHI Scores between the Audio-CLAP baseline and BTS. The results are shown depending on the metadata classes. Note that there is no sample of the LittC2SE in the test set. The bold characters and underlines indicate the \textbf{best} and \underline{second best} Score improvement.}
    
    \label{table4_metadata_anlysis}
    \renewcommand{\arraystretch}{1}
    \addtolength{\tabcolsep}{10pt}
    \resizebox{\linewidth}{!}{
    \begin{tabular}{llr|cc|c}
    \toprule
    %\multicolumn{1}{c}{\multirow{2}{*}{\begin{tabular}[c]{@{}c@{}}Difference\\ (BTS-Audio-CLAP)\end{tabular}}}
    \multicolumn{3}{c}{Metadata} & \multicolumn{2}{c}{Method} & Score \\

    Type & Class & Ratio (\%) & BTS & Audio-CLAP & Difference \\
    %\cmidrule(l{2pt}r{2pt}){4-6} 
    %& & & \multicolumn{3}{c}{Score (\%)} \\

    \hline \midrule
    \multirow{2}{*}{Age} & Adult & 85.70 & 64.53 & 61.67 & 2.86 \\
    & Pediatric & 14.30 & 64.53 & 61.99 & 2.54 \\
    
    \bottomrule
    \multirow{2}{*}{Sex} & Male & 78.74 & 64.53 & 62.00 & 2.53 \\
    & Female & 21.26 & 64.46 & 61.92 & 2.54 \\

    \bottomrule
    \multirow{7}{*}{Loc} & Trachea & 11.97 & 64.46 & 61.92 & 2.54 \\
    & Left Anterior & 21.99 & 64.52 & 61.66 & 2.86 \\
    & Right Anterior & 9.51 & 64.78 & 61.58 & 3.20 \\
    & Left Posterior & 22.57 & 64.54 & 62.00 & 2.54 \\
    & Right Posterior & 15.64 & 65.31 & 61.45 & \underline{3.86} \\
    & Left Lateral & 9.43 & 64.41 & 61.83 & 2.58 \\
    & Right Lateral & 8.89 & 60.21 & 54.44 & \textbf{5.77} \\
    
    \bottomrule
    \multirow{4}{*}{Dev} & Meditron & 16.65 & 64.54 & 62.00 & 2.54 \\
    & LittC2SE & 0.0 & - & - & - \\
    & Litt3200 & 16.73 & 64.52 & 61.65 & 2.87 \\
    & AKGC417L & 66.62 & 64.78 & 61.53 & 3.25 \\
    
    \bottomrule
    \end{tabular}}
    \vspace{-5mm}
\end{table}

Table~\ref{table3_main_results} presents comprehensive ICBHI results including our method.
% the experiment results of the proposed approach and baselines.
Our method achieves a new SOTA by 1.17\% improvement from the previous best without using any additional training techniques that other methods rely on, such as stethoscope-specific fine-tuning~\cite{Gairola21}, co-tuning~\cite{nguyen2022lung}, Patch-Mix augmentation~\cite{bae23b_interspeech}, or domain adaptation techniques~\cite{kim2023stethoscope, kim2023adversarial}.
Particularly, it is noteworthy that our method has a considerably higher sensitivity ($S_e$) than the previous best model while maintaining a similar specificity ($S_{p}$).
We consider that the additional context from textual descriptions enhances the model's ability to correctly identify positive cases, without increasing the false positive ratio.
Additionally, the improvement of Audio-CLAP over the previous SOTA highlights the effectiveness of the contrastively language-audio pretrained encoder as a strong baseline for audio-only related tasks, where the encoder is pretrained using text descriptions as opposed to previous backbone models for RSC that employed categorical audio labels for pretraining.
%Additionally, the improvement of Audio-CLAP over the previous SOTA highlights the effectiveness of the language-audio pretrained encoder as a strong baseline for audio-only related tasks. 

%This result suggests that utilizing ...

Table~\ref{table4_metadata_anlysis} compares the performance of BTS and Audio-CLAP for each metadata category within the ICBHI test set. 
Although the dataset is notably imbalanced for the metadata categories, our model consistently surpasses the Audio-CLAP baseline across all classes.
It is also noteworthy that a notable enhancement is observed in minority classes. Specifically, the result of the test samples, of which \emph{Loc} is the right lateral chest, yields the Score increase of 5.77\%. This underscores the value of the metadata in not only the overall performance improvement but also the effectiveness of accounting for underrepresented categories.
%This underscores the value of the metadata in not only the overall performance improvement but also the effectiveness on underrepresented categories.

\subsection{Influence of Metadata on Classification Performance}

\begin{table}[!t]
    \centering
    %\vspace{-3mm}
    \caption{Results of the ablation study with different combinations of the metadata. The \textbf{best} result is indicated by the bold.}
    %Performance gap according to various metadata combinations on the BTS method. Using all of the metadata achieves the best result.
    
    \label{table5_subset}
    \renewcommand{\arraystretch}{1}
    \addtolength{\tabcolsep}{8pt}
    \resizebox{\linewidth}{!}{
    \begin{tabular}{lcc|lll}
    \toprule
    Method & Setting & Metadata  & $S_p$\,(\%) & $S_e$\,(\%) & \textbf{Score}\,(\%) \\
    \hline \midrule

    \multirow{5}{*}{BTS} & (1) & All & $\text{81.40}_{\pm 2.57}$ & $\text{45.67}_{\pm 2.66}$ & $\text{\textbf{63.54}}_{\pm 0.80}$ \\
    
    & (2) & Age-Sex-Loc & $\text{81.71}_{\pm 4.12}$ & $\text{43.63}_{\pm 3.48}$ & $\text{62.66}_{\pm 0.35}$ \\

    & (2) & Age-Sex-Dev &  $\text{79.49}_{\pm 3.66}$ & $\text{46.04}_{\pm 2.39}$ & $\text{62.76}_{\pm 1.09}$ \\

    & (2) & Age-Loc-Dev  & $\text{82.28}_{\pm 5.27}$ & $\text{43.48}_{\pm 4.38}$ & $\text{62.88}_{\pm 0.83}$ \\

    & (2) & Sex-Loc-Dev & $\text{84.66}_{\pm 3.63}$ & $\text{41.09}_{\pm 3.37}$ & $\text{62.88}_{\pm 0.54}$ \\

    \midrule

    Audio-CLAP & (3) & - &  $\text{80.85}_{\pm 3.33}$ & $\text{44.67}_{\pm 3.77}$ & $\text{62.56}_{\pm 0.37}$ \\
    
    \bottomrule
    \end{tabular}}
    \vspace{-5mm}
\end{table}
We analyze the impact of metadata on classification performance by comparing the results of three distinct experiment settings: (1) the full set of metadata with BTS, (2) subset with exclusion of a single attribute with BTS, and (3) audio-encoder only in the case of Audio-CLAP. The results are summarized in Table~\ref{table5_subset}, which shows that more textual context results in higher performance. 
Specifically, using all metadata in (1) yields the highest Score, while audio-encoder only (3) scored the lowest. The results of all metadata subsets (2) fall in between them. 
%Specifically, using all metadata (1) yields the highest Score, while using none (3) scored the lowest. The results of all metadata subsets (2) fall in between them.

Furthermore, the results demonstrate that the measurement location (\emph{Loc}) and recording device type (\emph{Dev}) have a larger effect than the demographic attributes, i.e., age and gender of patients.
The absence of \emph{Loc} and \emph{Dev} leads \emph{Score} drop of 0.78\% and 0.88\%, respectively, compared to the all metadata case.
%The subsets that included both \emph{Loc} and \emph{Dev} resulted in the Score of 62.88\%, which is notably higher than the results of other subsets.
This suggests that the type of recording device and the measurement location significantly influence the acoustic properties of respiratory sounds. Therefore, the combined use of these metadata provides the meaningful context to understand the respiratory sounds.

% \subsection{Evaluation in Real-world Metadata Scenarios}

% Table~\ref{table4_subset} presents the performance across the subsets corresponding to each metadata category within the ICBHI test set. 
% The dataset has notable in-balances with respect to the categories. Despite these imbalances, our model consistently surpasses the Audio-CLAP baseline across all classes.
% Particularly noteworthy is the significant enhancement observed in minority classes, specifically in the case of Loc-Right Lateral, with an increase of as much as 5.77\%. This underscores the value of the metadata attributes in not only increasing the overall performance, but the effectiveness on underrepresented categories.

%\red{- Table 4 shows the \emph{Score} according to various metadata of the ICBHI test set}

%\red{- It looks very imbalanced (adult vs. pediatric, Male vs. Female, etc)}

%\red{- Our BTS outperforms the Audio-CLAP in theview of all metadata. }

%\red{- Emphasize about `Significantly improved in the minority cases: Loc-Right Anterior, Loc-Right Lateral)'}

%Subset experiments
%- Check which metadata contributes the performance improved
%- Full vs Full - age/sex/loc/dev (i.e., Removing one of the metadata from all
%If the loc and dev metadata is included together, the performance is increased.
%- This contributes more the good performance
%BTS-all is best
%- More information is better

\vspace{-2mm}

\subsection{Unknown Metadata Scenario}
%\subsection{Unknown Metadata Scenario: BMI}

To probe how well the model generalizes to unseen text descriptions, we examined the model performance for unseen test data, which additionally includes a new metadata attribute that is not used for training. Specifically, we added a sentence to describe the BMI of patients to the test data. The additional sentence is written in the same style as the training descriptions, e.g., ``The BMI of the patient was 20.50.''. The evaluation result is denoted as BTS[BMI] in Table~\ref{table6_bmi}.
% The evaluation result corresponds to settin as BTS[BMI] in Table~\ref{table6_bmi}.
Adding unknown metadata to text descriptions at test time shows only minor performance degradation, which suggests that the model performs reliably even with the unexpected metadata.

%Setting

%- We put the additional BMI description. E.g,  %`This sound was recorded from the left anterior chest of an adult male patient, using a Meditron stethoscope. The BMI of the patient was {20.50}.'.

%Results:
% - Classify reliably even when incoming metadata is different from what the model trained with

\subsection{Missing Metadata Scenarios}

\begin{table}[!t]
    \centering
    %\vspace{-3mm}
    %\caption{Results with additional metadata attribute BMI.}

    \caption{Results with variations on the metadata. The variations include the additional BMI attribute, partial metadata, and no metadata in text description. The \textbf{Best} result.}
    
    \label{table6_bmi}
    \renewcommand{\arraystretch}{1}
    \addtolength{\tabcolsep}{8pt}
    \resizebox{\linewidth}{!}{
    \begin{tabular}{l|clll}
    \toprule
    Method & $S_p$\,(\%) & $S_e$\,(\%) & \textbf{Score}\,(\%) \\
    \hline \midrule

    BTS & $\text{81.40}_{\pm 2.57}$ & $\text{45.67}_{\pm 2.66}$ & $\text{\textbf{63.54}}_{\pm 0.80}$ \\

    \midrule

    BTS[BMI] & $\text{81.40}_{\pm 2.57}$ & $\text{45.66}_{\pm 2.65}$ & $\text{63.53}_{\pm 0.80}$ \\

    \midrule

    BTS[Partial Metadata] & $\text{81.29}_{\pm 2.55}$ & $\text{45.54}_{\pm 2.61}$ & $\text{63.41}_{\pm 0.78}$ \\

    BTS[No Metadata] & $\text{80.82}_{\pm 3.54}$ & $\text{45.59}_{\pm 2.59}$ & $\text{63.21}_{\pm 1.04}$ \\
    
    %BTS[No Description] & $\text{80.82}_{\pm 3.54}$ & $\text{45.59}_{\pm 2.59}$ & $\text{63.21}_{\pm 1.04}$ \\

    %\bottomrule
    \midrule

    Audio-CLAP & $\text{80.85}_{\pm 3.33}$ & $\text{44.67}_{\pm 3.77}$ & $\text{62.56}_{\pm 0.37}$ \\

    \bottomrule
    \end{tabular}}
    \vspace{-5mm}
\end{table}
To understand how metadata that is partially or entirely missing effect the model, we conducted two experiments. 
%where the metadata is partially or entirely missing. 
We first partially removed the metadata (BTS[Partial Metadata]), by randomly eliminating one of the metadata attributes from test samples and substituting ``Unknown'' with 10\% probability. Then, we entirely removed the metadata (BTS[No Metadata]) and replace the whole description by ``No description.''
Table~\ref{table6_bmi} describes the experiment results. As expected, BTS[Partial Metadata] shows a slightly degraded Score compared to BTS, while BTS[No Metadata] results in a relatively large performance reduction. Nevertheless, the results with missing metadata maintain an edge over Audio-CLAP.
The results show that the BTS model is robust to missing metadata. We conjecture that the model learns to infer certain metadata characteristics directly from the audio, thereby preserving its strong performance even in the absence of metadata during inference.

%Our observations suggest that the classifier has learned a structural understanding of the audio records through the metadata provided, leading to consistent model performance even in the absence of expected text information during inference.
%We infer that the classifier learned to structurally understand the audio records based on the given metadata, which results in the reliable performance of the model even if the expected text information is not available at inference time.

% The integration of text descriptions contributes to a more stable training process, but becomes less useful for inference stage as the model learns. 
% this needs some chart
% Specifically, as training progresses, we notice a reduced dependency on text outputs, as the model begins to focus more intently on audio data.
%Outputs of two encoders are affected by each other. 
%- Text descriptions lead model training to more stable
%When training is going, dependency of text outputs are lower (i.e., model can be focused on audio)
%- Text audio encoder can be used as a good auxiliary module
%Audio is more important
%- There are various cycle in one waveform, but the metadata is same
%- Thus, because there are the same metadata in the training sets, model can be focus on audio

\section{Conclusion}
In this work, we proposed to directly utilize the metadata to improve the performance of RSC.
Our experiments demonstrated that including the metadata as additional context for the classification leads to a considerable performance increase, which results in the new SOTA for RSC.
In particular, the experiment results showed that our method helps minimize the performance degradation due to the acoustic variations induced by the inherent factors relating to the demographics and recording environment. Moreover, our method works reliably even when the metadata is with unexpected information, partially unavailable, or even completely unavailable.
Besides, we showed that the CLAP model provides a strong baseline for audio-only tasks.
%Besides, our results indicate that the CLAP model provides a strong baseline for audio encoder. 
%We believe that our approach can be expanded to various other text and sound multimodal tasks.
%future?
%We believe that our approach can be expanded to the various medical domain downstream tasks.
%We found that utilizing pretrained CLAP model not only provides a strong new baseline in audio-only setting, but also considerably outperforms the previous state-of-the-art when textual descriptions of the metadata are included during training and inference. 
% future work: analyze the impact of imbalances in the count of dev, loc, etc on the performance

\section{Acknowledgement}
This research was supported by the MSIT (Ministry of Science and ICT), Korea, under the ITRC (Information Technology Research Center) support program (IITP-2024-2020-0-01808) supervised by the IITP (Institute of Information \& Communications Technology Planning \& Evaluation) and by Brian Impact Foundation, a non-profit organization dedicated to the advancement of science and technology for all.

\bibliographystyle{IEEEtran}
\bibliography{mybib}

\end{document}